\documentclass[11pt,titlepage]{article}
\usepackage{graphicx}
\usepackage[english]{babel}
\usepackage{amsmath}
\usepackage{amssymb}
\usepackage{setspace}
\usepackage[round,numbers,sort&compress]{natbib}

\begin{document}
\title{The influence of the cylindrical shape of the nucleosomes and H1 defects on properties of chromatin}
\author{Philipp M. Diesinger and Dieter W. Heermann
         \\ {} \\
           Institut f\"ur theoretische Physik \\
           Universit\"at Heidelberg \\
           Philosophenweg 19 \\
           D-69120 Heidelberg\\
           and\\
           Interdisziplin\"ares Zentrum\\
           f\"ur Wissenschaftliches Rechnen\\
           der Universit\"at Heidelberg \\ {} \\
\vspace {3ex}}

\newpage

\author{Philipp M. Diesinger\thanks{
         Corresponding author.  Address:
         Institut f\"ur theoretische Physik,
         Philosophenweg 19,
         D-69120 Heidelberg,
         Germany,
         Tel.:~(+49)6221-54-9449, Fax:~(+49)-6221-549-331,
         email:~diesinger@tphys.uni-heidelberg.de
       } \\
    Department of Physics \\
    Institut f\"ur theoretische Physik \\Heidelberg University \\Germany
    \and Dieter W. Heermann \\ Department of Physics \\
    Institut f\"ur theoretische Physik \\ and \\
               Interdisziplin\"ares Zentrum\\
           f\"ur Wissenschaftliches Rechnen\\
    Heidelberg University, Germany}

\date{}

\maketitle \vfill\eject
\begin{abstract}
We present a model improving the two-angle model for interphase chromatin (E2A model). This model takes into account the cylindrical shape of
the histone octamers, the H1 histones in front of the nucleosomes and the vertical distance $d$ between the in and outgoing DNA strands.
Factoring these chromatin features in, one gets essential changes in the chromatin phase diagram: Not only the shape of the excluded-volume
borderline changes but also the vertical distance $d$ has a dramatic influence on the forbidden area. Furthermore, we examined the influence of
H1 defects on the properties of the chromatin fiber. Thus we present two possible strategies for chromatin compaction: The use of very dense
states in the phase diagram in the gaps in the excluded volume borderline or missing H1 histones which can lead to very compact fibers. The
chromatin fiber might use both of these mechanisms to compact itself at least locally. Line densities computed within the model coincident with
the experimental values.
\end{abstract}

\vspace{2cm} {\bf Keywords: chromatin, phase diagram, cylindrical
nucleosomes, histone H1, excluded volume interaction, Monte Carlo
method}

\newpage
\section*{Introduction}
The nucleosome is the basic repeat unit of chromatin~\cite{van Holde} in all eucaryotic organisms. It consists of a cylindrical-shaped histone
octamer and a stretch of DNA which is wrapped around the histone complex approximately 1.65 times. The histone octamer consists of four pairs of
core histones (H2A, H2B, H3 and H4) and is known up to atomistic resolution~\cite{Davey,Luger}. The nucleosomes are connected by blank DNA
strands and together with these linkers they form the so-called 30nm fiber. The histone H1 (and the variant histone H5 with similar structure
and functions) is involved in the packing of the beads on a string structure into the 30nm chromatin structure. To do so it sits in front of the
nucleosome keeping in place the DNA which is wrapped around the histone octamer and thus stabilizes the chromatin fiber. The nucleosome provides
the lowest level of compaction and, furthermore, it is important in the regulation of transcription. Several enzymes can change the position of
the nucleosome~\cite{Lia} to make the genetic information which is held within the nucleosome core particle accessible.

$\\$ The compaction of the DNA plays a very important role in modern biophysics since it has a total length of some meters but has to fit into a
nucleus of some microns. The degree of compaction depends on the salt concentration~\cite{Thoma} and on the presence of linker
histones~\cite{Bednar}. The presence of the linker histones leads to the formation of stem-like structures which are formed by the incoming and
outcoming DNA string in front of the nucleosome. At low salt concentration a 10nm structure is formed which has the shape of beads on a
string~\cite{Thoma} whereas at high salt concentrations the chromatin fiber is much more compact and has a diameter of 30nm~\cite{Widom86}.

$\\$The chromatin structure is still not completely understood~\cite{Chakravarthy,van Holde,Zlatanova,Zlatanova2}. There are competing models
for its structure: zigzag ribbon models~\cite{Schalch,Bednar,Woodcock,Schiessel2,Dorigo}, helical solenoid models~\cite{Thoma,Finch,Widom10},
superbeads or simply having no regular structure~\cite{van_Holde2}. A crystal structure of a tetranucleosome has been revealed~\cite{Schalch}
and used to construct a model for the 30nm fiber which resembles a zigzag ribbon that twists or supercoils. The chromatin fiber has been
investigated by electron cryo-microscopy~\cite{Bednar,Bednar98}, atomic force microscopy~\cite{Leuba,Zlatanova3} neutron scattering and scanning
transmission electron microscopy~\cite{Gerchman}. Beyond the 30nm level chromatin is poorly understood.

$\\$The two-angle model was introduced by Woodcock et al.~\cite{Woodcock} to describe the geometry of the 30nm chromatin fiber. It has been
shown that the excluded volume of the histone complex plays a very important role for the stiffness of the chromatin fiber~\cite{Mergell} and
for the topological constraints during condensation/decondensation processes~\cite{Barbi}. In~\cite{Schiessel} a rough approximation of the
forbidden surface in the chromatin phase diagram was given. In a previous work of ours~\cite{Diesinger} we answered questions concerning the
fine structure of the excluded volume borderline which separates the allowed and forbidden states in the phase diagram with the basic assumption
of spherical nucleosomes and no vertical shift between in and outgoing strand. Here we present a Ramachandran-like diagram for chromatin fibers
with cylindrical nucleosomes for a new extended model and furthermore discuss the influence of a vertical shift between the linkers due to H1
histones and the volume exclusion of the DNA.

We first present the basic notations for the formulation of the
Extended Two-Angle model (E2A). Then we give an algorithm for the
generation of chains within the model and present the resulting
phase diagram and end-to-end distance as well as radius of
gyration results.

\section*{Theory}
\subsection*{Extended two-angle model}
We extend the two-angle model by introducing a parameter for the vertical distance between the DNA strands in front of the nucleosome.
Furthermore, we take the cylindrical excluded volume of the nucleosomes into account as well as the H1 histones which fix the DNA linkers in
front of the nucleosome. The H1 histones themselves are taken to be random variables to allow for possible missing H1 histones.

\subsection*{Basic notations}
We start out and fix some basic notations to use for the formulation of the model. The nucleosomes will be characterized by the centers $N_i \in
\mathbb{R}^3$ and the orientations $\hat{p}_i \in \mathbb{R}^3$ of the nucleosomes, with $i=0,...,N$ and $\|\hat{p}_i\|=1$. $N$ is the length of
the fiber. The linkers between the centers of two nucleosomes will be denoted by $b_i := N_i - N_{i-1}$ with $i=1,...,N$. The length $\| b_i \|$
of the linkers will be a further input parameter of the model (opposite of the \emph{direction} $b_i \in \mathbb{R}^3$ of the linkers).
Furthermore, the entry-exit-angle $\alpha_i \in [0,\pi]$ between two consecutive linkers is defined by $\alpha_i := \sphericalangle
(-b_i,b_{i+1})$ with $i=1,...,N-1$ and the rotational angle $\beta_i \in [0,\pi]$ between two consecutive orientations is given by $\beta_i :=
\sphericalangle (p_{i-1},p_i)$ with $ i=1,...,N$.

Moreover, $h_i$ represents the distance along the orientational axis $\hat{p}_{i-1}$ from $N_{i-1}$ to $N_i$ due to the spatial discrepancy
between in and outgoing DNA strand. $h_i$ can be expressed by the vertical distances $d_i$ which the DNA covers by wrapping up itself around the
histone complexes: $h_i = \frac{1}{2}(d_{i-1}+d_i)$ with $i=1,...,N$.

Thus a single chromatin strand within the two-angle-model is characterized by the following set of variables
\begin{equation*}
(\{\alpha_i\}_{i \in \{1,..,N-1\}},\{\beta_i\}_{i \in
\{1,..,N\}},\{ h_i\}_{i \in \{1,..,N\}},\{\|b_i\|\}_{i \in
\{1,..,N\}}).
\end{equation*}

The general rotational matrix $\mathcal{R}$ around an axis $\hat{a} = (a_1,a_2,a_3)^t$ (with $\|\hat{a}\|=1$) by an angle $\gamma$ with respect
to the right-hand rule will be used in the following. It is given by:
\begin{equation*}\tiny \mathcal{R}_{\hat{a}}^{\gamma} = \left( \begin{array}{ccc} \text{cos}\gamma+a_1^2(1-\text{cos}\gamma) & a_1 a_2 (1-\text{cos}\gamma)- a_3 \text{sin} \gamma & a_1 a_3 (1-\text{cos}\gamma) + a_2 \text{sin} \gamma \\
a_2a_1(1-\text{cos} \gamma) +a_3 \text{sin} \gamma & \text{cos} \gamma +v_2^2 (1 - \text{cos} \gamma)& a_2 a_3 (1-\text{cos}\gamma) -a_1 \text{sin} \gamma \\
a_3a_1(1-\text{cos} \gamma) - a_2 \text{sin} \gamma & a_3 a_2 (1-\text{cos} \gamma) + a_1 \text{sin} \gamma & \text{cos} \gamma + a_3^2
(1-\text{cos} \gamma)
\end{array} \right ).
\end{equation*}

\subsection*{Definition of the two-angle model}
\begin{figure}[H]\begin{center}
\includegraphics[width=\textwidth ,angle=0]{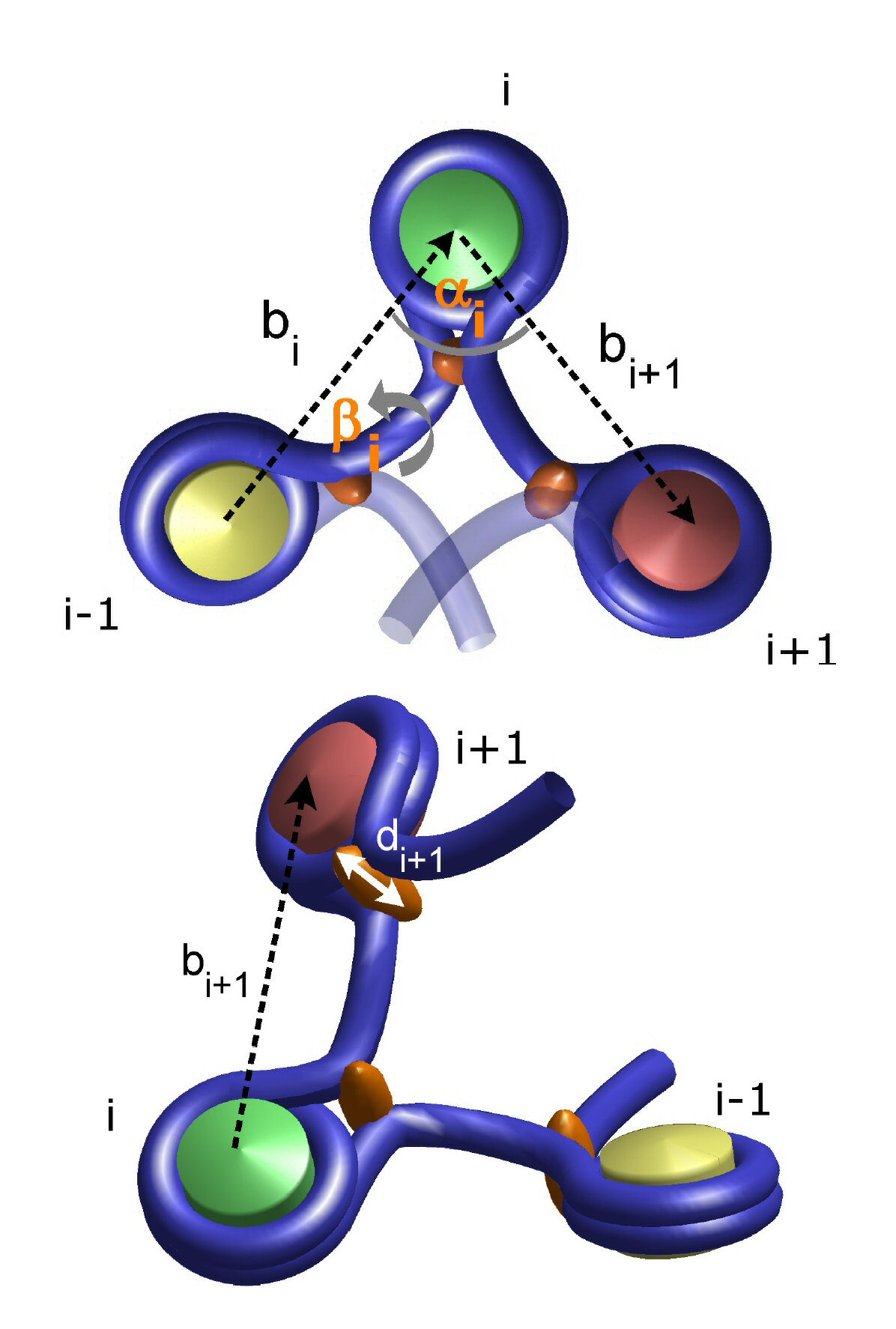}
\caption{\label{fig:basic_def} The figure shows the basic parameters of the E2A model: The entry-exit-angle $\alpha_i$, the rotational angle
$\beta_i$, the linker length $b_i$ and the vertical distance $d_i$ between in and outgoing linker. We chose a large entry-exit-angle here to
make the visualization clear.}
\end{center}\end{figure}

A chromatin fiber within the framework of the extended two-angle model has to fulfil the following conditions for all $i \in \{1,...,N\}$:
\begin{align*}
&i) \;\;\;\sphericalangle \left ( -b_i, b_{i+1} \right ) = \alpha_i \Leftrightarrow \text{\text{cos}} (\alpha_i) = \frac{\langle -b_i, b_{i+1} \rangle}{\|b_i\| \|b_{i+1}\|} \\
&ii)\;\; \sphericalangle(\hat{p}_{i-1},\hat{p}_{i})=\beta_i \Leftrightarrow \text{\text{cos}} (\beta_i) = \langle \hat{p}_{i-1}, \hat{p}_i \rangle \\
&iii)\;\;  \| N_i-N_{i-1}\| = \| b_i \| \\
&iv) \;\; \langle  \hat{p}_{i-1}, b_i \rangle \hat{p}_{i-1} = h_i
= \frac{1}{2}(d_{i-1}+d_i).
\end{align*}
These are illustrated in Fig~\ref{fig:basic_def}.

The first condition adjusts the entry-exit angle of nucleosome $i$
to the given parameter $\alpha_i$. The second condition does the
same for the rotational angle due to the DNA twist from nucleosome
$i-1$ to nucleosome $i$. The third condition fixes the distance of
the two consecutive nucleosomes $i-1$ and $i$ and the last
condition adjusts the vertical distance along the local chromatin
axis between the nucleosomes $i-1$ and $i$.

\subsection*{Construction of the fiber}
The construction of the fiber can be done using an iterative process. A further part of the model is the presence of a H1 histone which is
assumed to be present with probability $p$.

The first nucleosome center and its orientation are arbitrary. We
chose:
\begin{equation*} N_0 = \left( \begin{array}{c} 0 \\ 0 \\ 0\end{array} \right ), \quad
                  \hat{p}_0 = \left( \begin{array}{c} 0 \\ 0 \\ 1\end{array} \right ).
\end{equation*}

The following vectors fulfil the conditions of the two angle model for the second nucleosome location and its orientation:
\begin{equation*} N_1 = N_0 + \sqrt{ \|b_1\|^2-h_1^2} \left( \begin{array}{c} -1 \\ 0 \\ 0 \end{array} \right) + h_1 \hat{p}_0
                 = \left( \begin{array}{c} -\sqrt{ \|b_1\|^2-h_1^2} \\ 0 \\ h_1 \end{array} \right )\end{equation*}
                 and
\begin{equation*} \hat{p}_1 = \mathcal{R}_{\hat{a}}^{\beta_1} \hat{p}_{i-1}\quad  \text{with} \quad \hat{a}=(1,0,0)^t.
\end{equation*}

Now we can calculate $N_{i+1}$ and $\hat{p}_{i+1}$ in dependence of $N_i$, $N_{i-1}$, $\hat{p}_i$ and $\hat{p}_{i-1}$. With
\begin{equation*} v_i := -b_i + \langle \hat{p}_i, -b_i \rangle \hat{p}_i
\end{equation*}
and
\begin{equation}\label{EQ:first_use_alpha_0} v_i' := \mathcal{R}_{\hat{p}_i}^{\alpha_0} \sqrt{b_{i+1}^2-d_{i+1}^2} \left ( \frac{v_i}{\|v_i\|} \right ) + h_{i+1}\hat{p}_i
\end{equation}
one gets the location of nucleosome $i+1$ by
\begin{equation*}N_{i+1}=N_i +v_i'.
\end{equation*}

$\alpha_0$ is the angle between the projections of $b_{i+1}$ and $-b_{i-1}$ onto an arbitrary plane orthogonal to $\hat{p}_i$. We need to
calculate the dependence of this projected entry-exit-angle $\alpha_0$ on the actual entry-exit-angle $\alpha$.

Note that $\alpha_0$ was used as entry-exit-angle in some other
publications~\cite{Schiessel2,Schiessel} but in this work it
denotes only the projection of the real entry-exit-angle $\alpha$.

Using the law of cosine one gets
\begin{equation}\label{EQ:l_1}
l^2 = b_i^2 + b_{i+1}^2 - 2 b_i b_{i+1} \text{cos}(\alpha).
\end{equation}
Now we will use an affine transformation $T$ to a new coordinate system $(x,y,z) \overset{T}{\rightarrow} (x',y',z')$ in order to get a second
relation for $l$. We shift the origin to $N_i$ and rotate our old coordinate system so that $\hat{p}_i$ corresponds to the new $z$-axis.
Furthermore,  the new $x$-axis has to coincide with the projection of $-b_i$ onto any plane orthogonal to $\hat{p}_i$. $\\$Obviously,
\begin{equation*}
l^2 = \| b_i + v_i' \|^2 = \| b_i' + v_i'' \|^2
\end{equation*}
with
\begin{equation*} b_i \overset{T}{\rightarrow} b_i' = \left( \begin{array}{c} \sqrt{b_i^2 - \langle \hat{p}_i, -b_i\rangle^2} \\ 0 \\ \langle \hat{p}_i, -b_i\rangle\end{array} \right )
\end{equation*}
and
\begin{equation*} v_i' \overset{T}{\rightarrow} v_i''= \left( \begin{array}{c} \text{cos}(\alpha_0)\sqrt{b_{i+1}^2-h_{i+1}^2} \\ \sqrt{b_{i+1}^2 - h_{i+1}^2 - \left(\text{cos}(\alpha_0)\sqrt{b_{i+1}^2-h_{i+1}^2} \, \right)^2 } \\ h_{i+1} \end{array} \right
).
\end{equation*}
This leads to
\begin{equation}\label{Eq:l_2}
l^2=b_{i+1}^2+b_i^2-2h_{i+1}  \langle \hat{p}_i, -b_i\rangle -2\text{cos}(\alpha_0)\sqrt{b_i^2- \langle \hat{p}_i,
-b_i\rangle^2}\sqrt{b_{i+1}^2-h_{i+1}^2}.
\end{equation}
By comparing Eq.\ref{EQ:l_1} and Eq.\ref{Eq:l_2} one gets eventually
\begin{equation*}
\text{cos}(\alpha_0)=\frac{b_i b_{i+1} \text{cos}(\alpha) - h_{i+1} \langle \hat{p}_i,-b_i\rangle}{\sqrt{b_{i+1}^2-h_{i+1}^2}\sqrt{b_i^2-\langle
\hat{p}_i,-b_i\rangle^2}}
\end{equation*}
with the boundary condition
\begin{equation}\label{Eq:alpha_min}
\alpha_0 > \alpha_{min} = \text{acos} \left ( \frac{(h_{i+1} + \|\langle \hat{p}_i,b_i \rangle \|)^2-b_{i+1}^2-b_i^2}{-2b_i b_{i+1}} \right )
\end{equation}
due to non-vanishing $d_i$ and $d_{i+1}$. The calculation of $N_{i+1}$ is complete, since we now know the dependence of $\alpha_0$ on $\alpha$
and therefore one can use Eq.\ref{EQ:first_use_alpha_0} to determine $N_{i+1}$. But one still has to calculate the orientation $p_{i+1}$ of
nucleosome $N_{i+1}$. Due to the fixation of the in and outgoing DNA strand by the H1 histones this orientation can be calculated by a rotation
around the following normalized axis $\hat{a}$:
\begin{equation*}
\hat{a}:= \frac{-b_{i+1}-\langle p_i,-b_{i+1}\rangle \hat{p_i}}{\|\hat{a}\|}.
\end{equation*}
$\hat{p}_{i+1}$ then follows by a rotation of $\hat{p}_{i}$ around this axis:
\begin{equation*}
\hat{p}_{i+1}=\mathcal{R}_{\hat{a}}^{ \beta_{i+1} } \hat{p}_i.
\end{equation*}

\section*{Methods}
\subsection*{Chromatin phase diagram}
\begin{figure}[H]\begin{center}
\includegraphics[width=\textwidth ,angle=0]{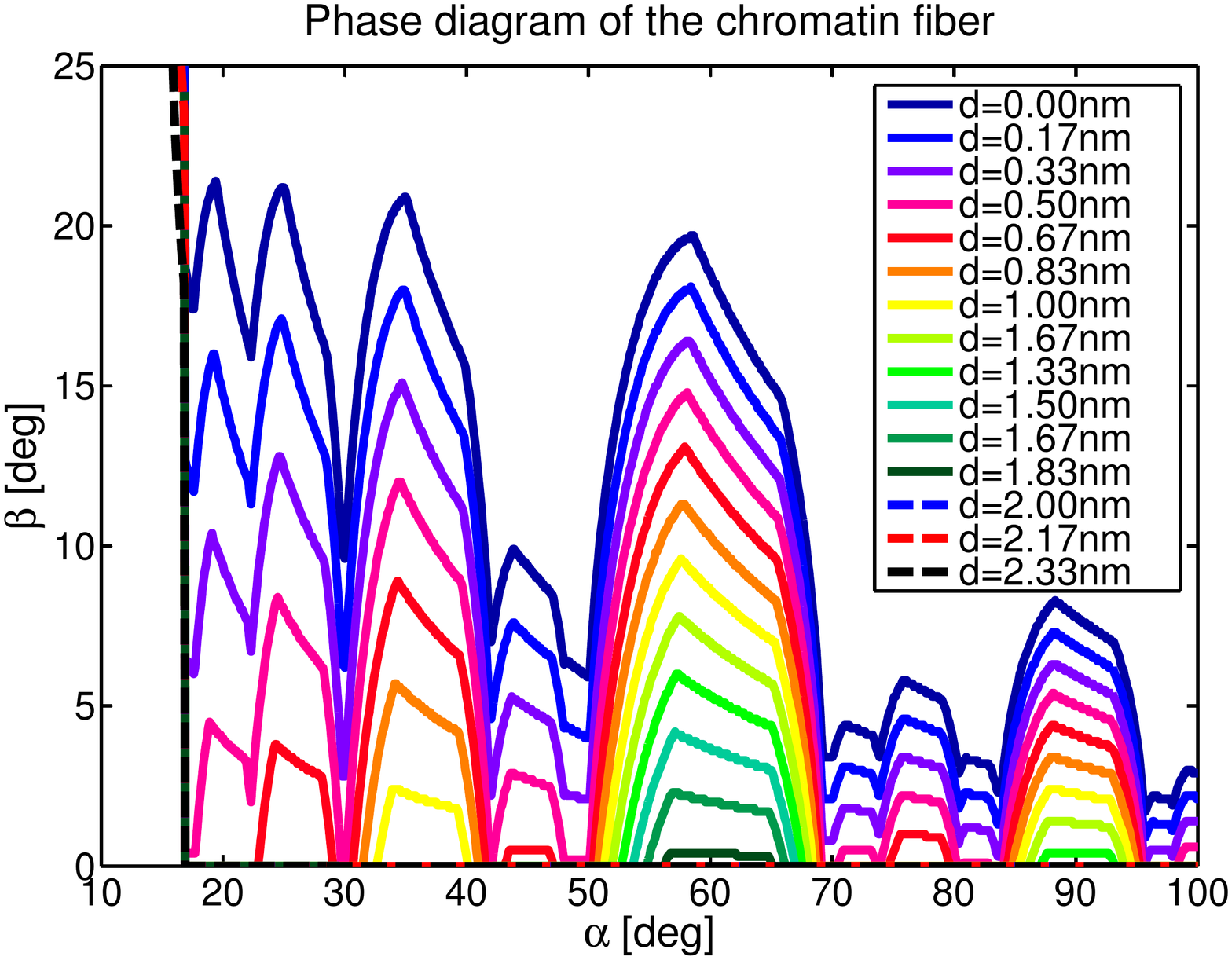}
\caption{\label{fig:phase_diagram1} Cut-out of the chromatin phase diagram (for different $d$). The states below the corresponding lines are
forbidden due to excluded volume interactions. With increasing $d$ more and more states become accessible to the fiber.}
\end{center}\end{figure}

First we determine the influence of the cylindrical excluded volume of the nucleosomes and a non-vanishing vertical distance between in and
outgoing DNA strand on the phase diagram of chromatin. Both of these parameters have been neglected so far \cite{Schiessel,Diesinger}. To do so
we made simulations~\cite{Binder} of regular chromatin fibers and checked whether they fulfil the excluded volume conditions or not. We were
able to plot our results in a Ramachandran-like diagram (cf.\ Fig.~\ref{fig:phase_diagram1}) and thus find out which states of the whole phase
diagram are forbidden by excluded volume interactions and which are not. The fibers we simulated for this part were \emph{regular}, i.e. all
linker lengths, entry-exit-angles, rotational angles and $h_i$ were fixed for a certain strand.

The cylinders had a height of 16.8$\langle \text{bp} \rangle$, and a diameter of 33.0$\langle \text{bp} \rangle$ according to \cite{Wolffe}.
They were orientated by using the vectors $p_i$ above. Moreover, we assumed a DNA diameter of 6.6$\langle \text{bp} \rangle$, a twist length of
10.2$\langle \text{bp} \rangle$, a mean linker length of 63$\langle \text{bp} \rangle$ and that 1.65 turns of DNA are wrapped around the histone
octameres.
\subsection*{Fibers with H1 defects}
Furthermore, we made Monte Carlo simulations of chromatin fibers with H1 defects, i.e. some of he H1 histones were missing. We used the
two-angle-model with some fixed parameters (see above), which reflect the probable mean values within the cell. The only interaction potential
is the hard core excluded volume. In this mean field-like approach we neglect~\cite{de_Gennes} thermal fluctuations and thus assume to be above
the $\Theta$-point to concentrate on the interaction between H1 defects and volume exclusion.

\begin{figure}[H]\begin{center}
\includegraphics[width=\textwidth ,angle=0]{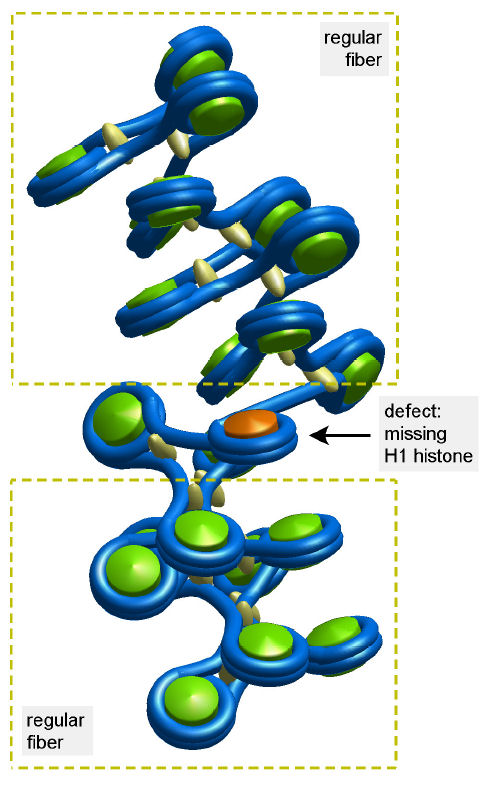}
\caption{\label{fig:Beispiel_H1_Fehlstelle} Shown is an example of a H1 defect within the chromatin fiber. The upper strand and the strand below
the defect are regular.}
\end{center}\end{figure}

For a certain nucleosome $N_i$ the defect probability $p$ gives
the chance of a missing H1 histone. If the histone is missing the
in and outgoing DNA strand are no longer fixed in front of the
nucleosome but instead are arbitrary with respect to the excluded
volume interactions of the chromatin strand (cf.\
Fig.~\ref{fig:Beispiel_H1_Fehlstelle}). Thus we get results for
the mean end-to-end distance and the mean radius of gyration of
fibers with various defect probabilities: $p=0.00$, $p=0.01$,
$p=0.05$, $p=0.10$ and $p=0.30$. For these simulations we fixed
the entry-exit-angles $\alpha_i$ to 40 degrees, the rotational
angles $\beta_i$ to 36 degrees and $h_i=7\langle \text{bp}
\rangle$.

\section*{Results}
\subsection*{Phase diagram}
\begin{figure}[H]\begin{center}
\includegraphics[width=.8\textwidth ,angle=0]{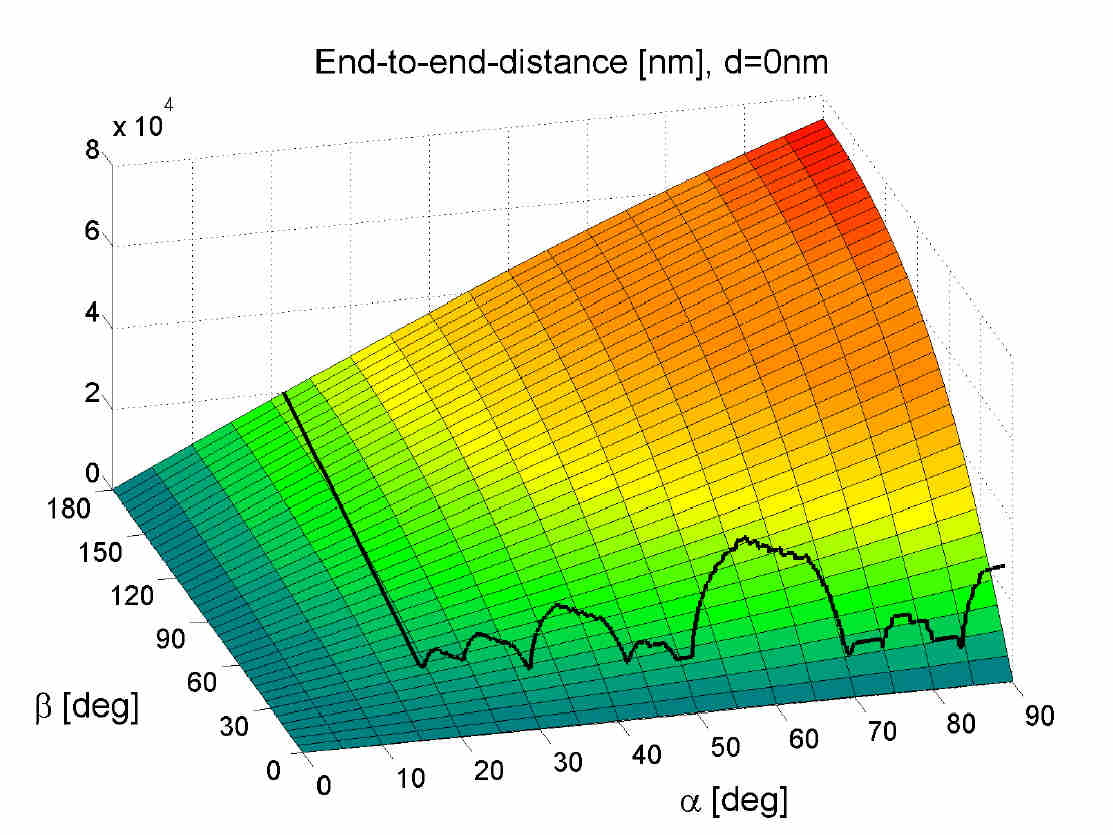}
\caption{\label{fig:end-to-end-distance_single_mesh2} The end-to-end distance of regular chromatin fibers along a cut-out of the phase diagram
with fixed fiber length ($N$=500 segments) and fixed $d$=0.0$\langle$bp$\rangle$. The solid black line represents the corresponding phase
transition.}
\end{center}\end{figure}

The colored lines in Fig.~\ref{fig:phase_diagram1} represent the phase transition between allowed and forbidden states. All states below the
corresponding line are forbidden, those above it are allowed. The states near the excluded volume borderline are the most interesting of the
phase diagram since they are the most compact ones (cf.\ Fig.~\ref{fig:end-to-end-distance_single_mesh2}). Therefore, the gaps in the borderline
might be used by the fiber to become (at least locally) very dense.

\begin{figure}[H]\begin{center}
\includegraphics[width=\textwidth ,angle=0]{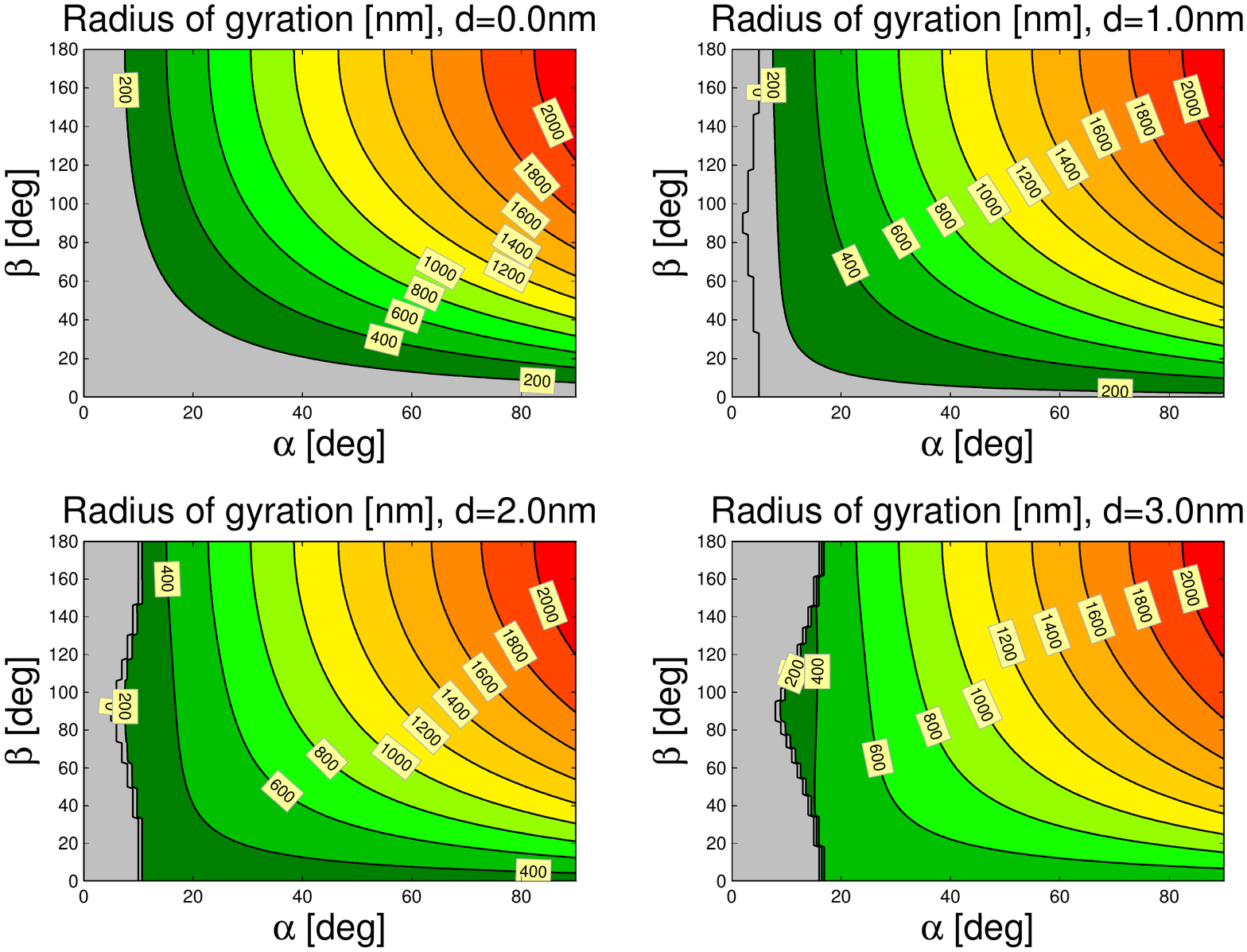}
\caption{\label{fig:radius_of_gyration2} The radius of gyration of regular chromatin fibers along a cut-out of the phase diagram with fixed
fiber length of $N$=500 segments. The compaction of the fibers decreases strongly with increasing $d$.}
\end{center}\end{figure}

There is another borderline at the left side of the diagram which prevents $\alpha$ from getting smaller than some minimal value
$\alpha_{min}(h)$ which depends on $h$. This arrow-like structure can be seen best in Fig.~\ref{fig:radius_of_gyration2}. It shifts towards
larger $\alpha_{min}(h)$ with increasing $d$. The gap in this line is a further consequence of the \emph{cylindrical} excluded volume and can
not be seen in the phase diagram for spherical nucleosomes \cite{Diesinger}.

As a consequence of the cylindrical instead of spherical excluded volume of the nucleosomes the shape of the peaks in the phase diagram is
changed: Their top shows a wedge-like shape due to the edges of the cylinders. With increasing entry-exit-angle $\alpha$ there is more space
between the nucleosomes which leads to a larger variety of allowed rotational angles $\beta$ and thus to the missing tip at the top of the
peaks. This effect gets weaker with increasing $d$: The edges which cut the peaks become more parallel to the $\alpha$-axis.

With increasing vertical distance between in and outgoing DNA strand there is also more space between consecutive nucleosomes which leads to a
decrease of the borderline. More and more states become accessible and with $h=5.5\langle \text{bp} \rangle$ the borderline almost vanishes. The
natural mean value of $h$ is approximately $h=2.8nm=8.4\langle \text{bp} \rangle$ due to 1.65 turns of DNA with a diameter of $2.2nm$. Lower $h$
values might occur where the DNA has less turns around the histone complex.

\begin{figure}[H]\begin{center}
\includegraphics[width=\textwidth ,angle=0]{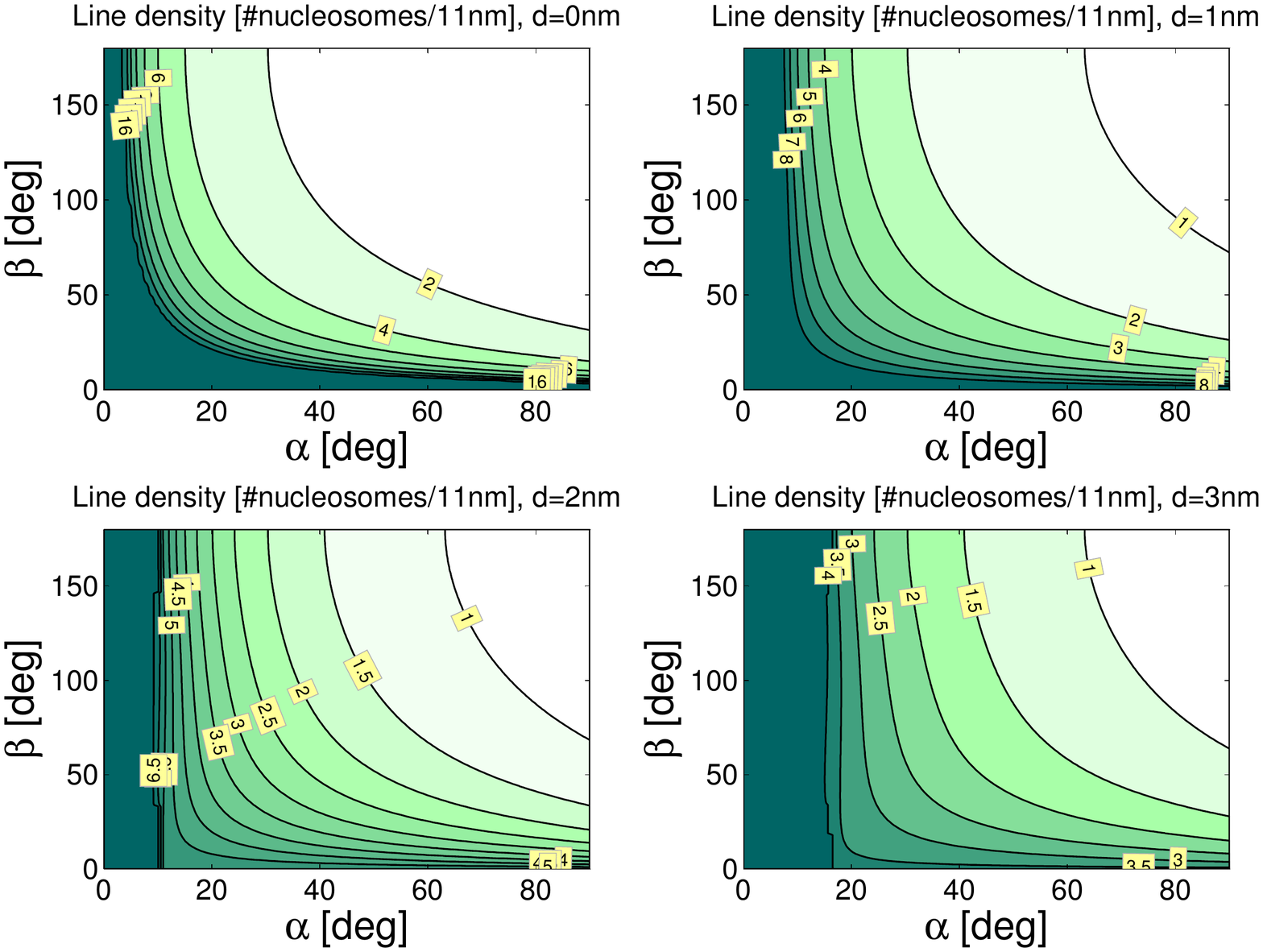}
\caption{\label{fig:line_density_chromatin2} The line density of the nucleosomes within a regular chromatin fiber is large for small $\beta_i$.
However the comparison with the phase diagram shows, that some of these states are forbidden due to excluded volume interactions.}
\end{center}\end{figure}

We furthermore examined the line density (cf.\ Fig.~\ref{fig:line_density_chromatin2}) and the radius of gyration (cf.\
Fig.~\ref{fig:radius_of_gyration2}) of regular chromatin fibers (length 500 nucleosomes) along the phase diagram. The most compact states can be
found near the excluded volume borderline. The line densities we found in our simulations coincide with experimental values~\cite{line_density}.
Increasing $d$ decreases the line density and increases the radius of gyration.

\subsection*{H1 defects}
\begin{figure}[H]\begin{center}
\includegraphics[width=.8\textwidth ,angle=0]{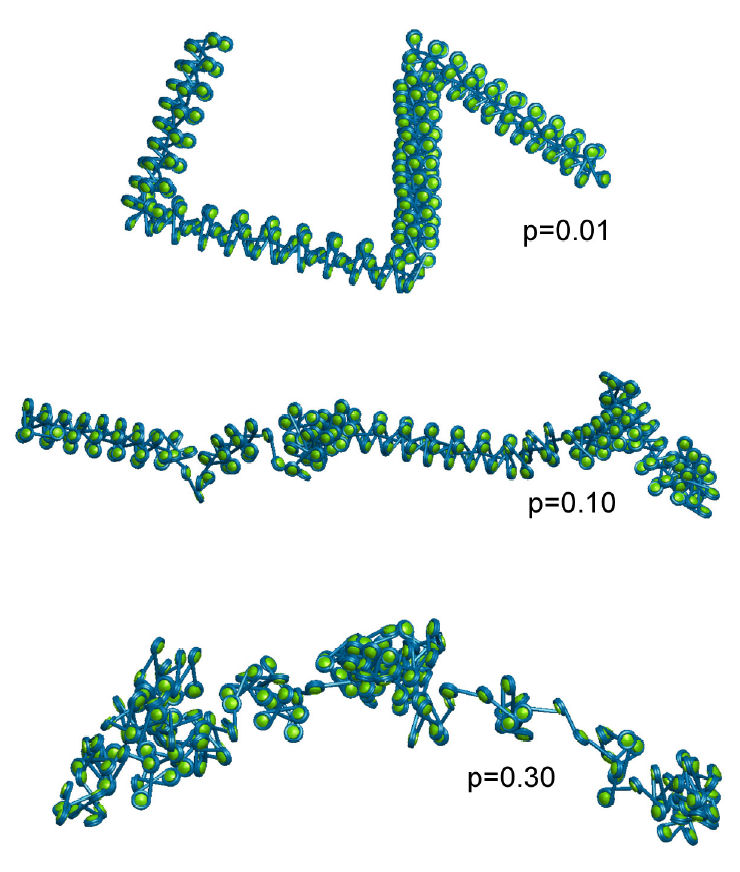}
\caption{\label{fig:fibervergleich} Chromatin fibers with different defect probabilities $p$. At $p=0.30$ the regular structure of the 30nm
strand is almost completely collapsed.  }
\end{center}\end{figure}

\begin{figure}[H]\begin{center}
\includegraphics[width=\textwidth ,angle=0]{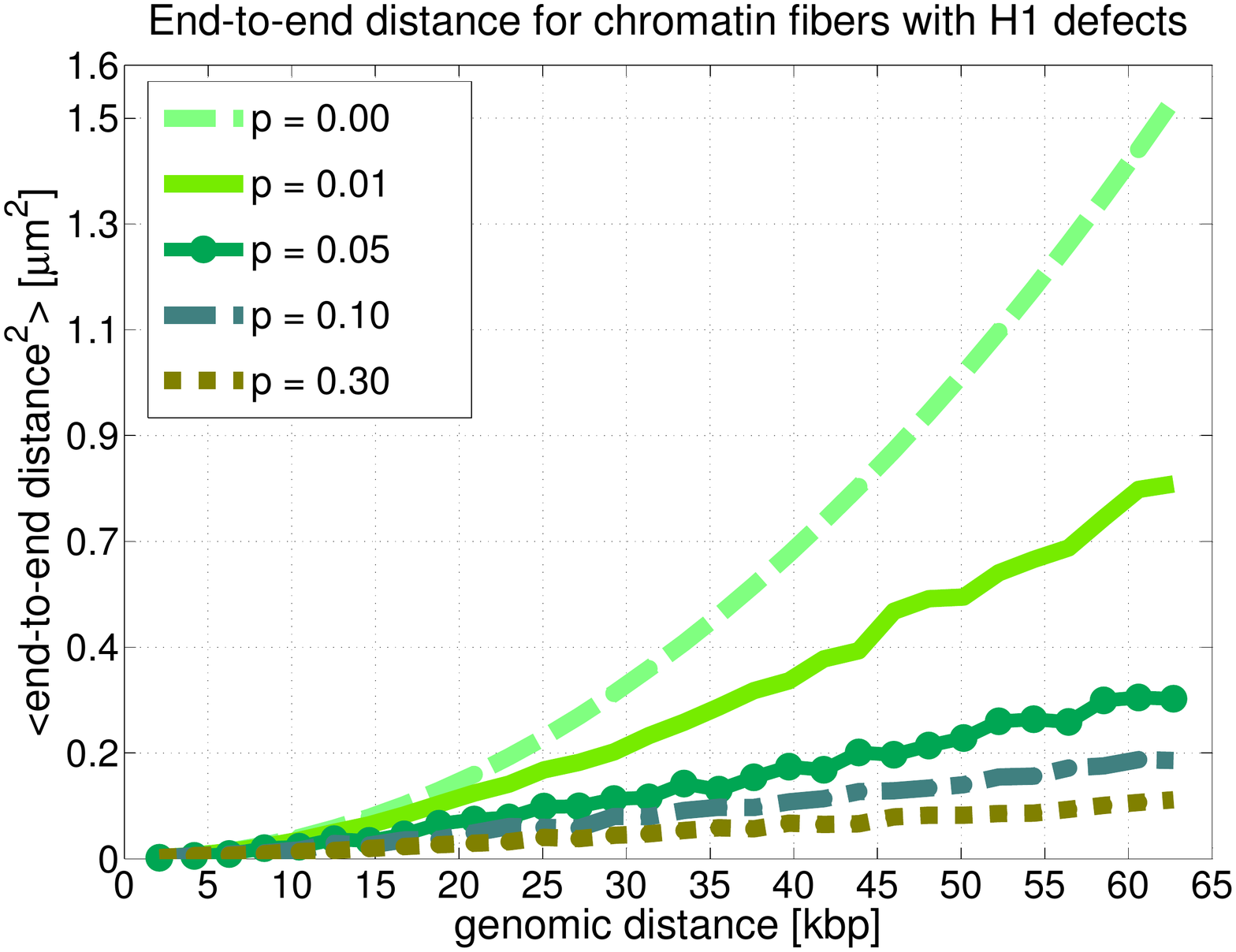}
\caption{\label{fig:Fehlstellen_EtE} The mean end-to-end distance for chromatin fibers with H1 defects. With
increasing defect probability $p$ the length of the fibers decreases rapidly. H1 defects might play a crucial
role for chromatin compaction.}
\end{center}\end{figure}

\begin{figure}[H]\begin{center}
\includegraphics[width=\textwidth ,angle=0]{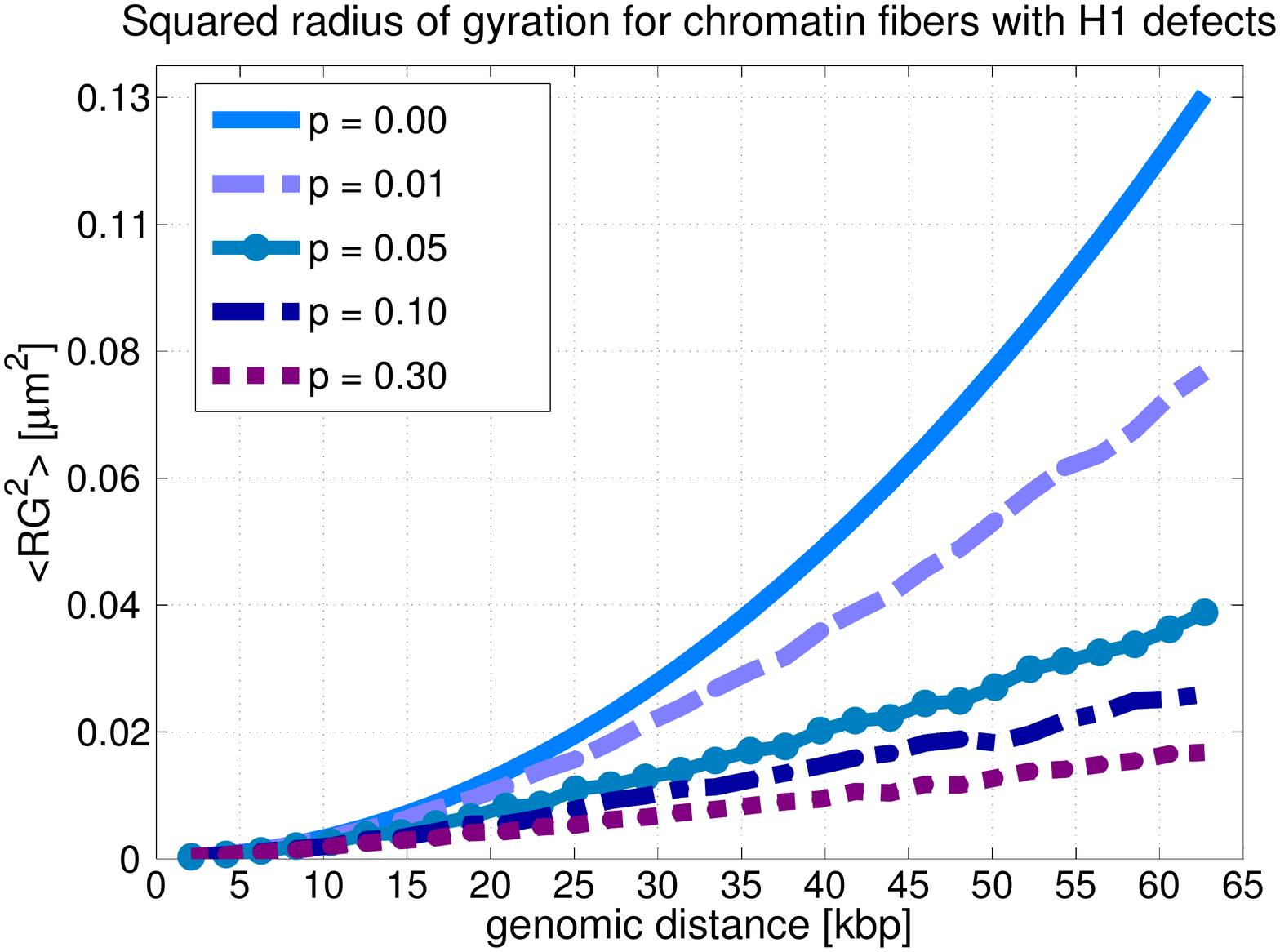}
\caption{\label{fig:Fehlstellen_RG} The squared radius of gyration for chromatin fibers with different defect
probabilities $p$. With increasing number of H1 defects the fiber becomes much more compact which could be an
important mechanism to compact the chromatin fiber.}
\end{center}\end{figure}

We also investigated the influence of missing H1 histones on the mean squared end-to-end distance (Fig.~\ref{fig:Fehlstellen_EtE}) and the mean
squared radius of gyration (Fig.~\ref{fig:Fehlstellen_RG}) of chromatin fibers. The parameter $p$ gives the defect probability in this section.
One can clearly see that even very small defect rates of some percent have a huge effect on the compaction of chromatin: Both mean squared
radius of gyration and the mean end-to-end distance decrease rapidly if one allows only a few H1 defects. Without H1 defects ($p=0$) we get an
ideal chromatin fiber within the restriction of the extended two-angle model. This ideal fiber reflects the properties of the 30nm strand only
on small length scales. Therefore, the increase of the compaction due to defects will probably be not as strong as implicated by our results.
Nevertheless missing H1 histones might contribute to chromatin compaction and DNA accessibility for transcription purposes at the same time
since one can see from Fig.~\ref{fig:fibervergleich} that although the fiber gets compacter some very open parts appear. One can also see here
the increasing disorder with increasing $p$.

\section*{Discussion}
The compaction of chromatin is still an open question. A polymer of a total length of two meters has to fit into a tiny cell nucleus of some
microns. We showed two possible strategies for the fiber to deal with this task. The fiber might use gaps in the phase diagram, i.e. very dense
states to compact parts of itself. Furthermore, we showed that missing H1 histones might supply a further contribution to the compaction of the
fiber. These H1 defects might play a very crucial role in the task of chromatin compaction and at the same time serve the transcription of the
DNA by opening locally the chromatin fiber. $\\$ Moreover, we developed the ordinary two-angle model further to our E2A model, which is much
more detailed and thus appropriate to model chromatin at the 30nm level.

\large $\\$Acknowledgements $\\\\$ \normalsize We thank Giacomo
Cavalli, J\"org Langowski and Roel van Driel for fruitful
discussions.

\vfill\eject

\vfill\eject

\end{document}